\documentclass[conference,a4paper]{APSIPA2017}
\usepackage{multirow}
\usepackage{graphicx}
\usepackage{amsmath}
\usepackage{subfigure}
\usepackage[psamsfonts]{amssymb}
\usepackage{amsxtra}
\usepackage{threeparttable}

\begin{document}

\title{Speaker Recognition with Cough, Laugh and ``Wei''}

\author{%
\authorblockN{%
Miao Zhang\authorrefmark{1}\authorrefmark{2},
Yixiang Chen\authorrefmark{1},
Lantian Li\authorrefmark{1} and
Dong Wang\authorrefmark{1}
}
\authorblockA{%
\authorrefmark{1}
Center for Speech and Language Technologies (CSLT), RIIT, Tsinghua University \\
Tsinghua National Laboratory for Information Science and Technology
}
\authorblockA{%
\authorrefmark{2}
Beijing University of Posts and Telecommunications
}
}

\maketitle
\thispagestyle{empty}

\makeatletter{\renewcommand*{\@makefnmark}{}
\footnotetext{M.Z. and Y.C. are joint first authors with equal contribution. }\makeatother}

\begin{abstract}

This paper proposes a speaker recognition (SRE) task with trivial speech events, such as cough and laugh.
These trivial events are ubiquitous in conversations and less subjected to
intentional change, therefore offering valuable particularities to discover the genuine speaker from disguised speech.
However, trivial events are often short and idiocratic in spectral patterns, making SRE extremely difficult.
Fortunately, we found a very powerful deep feature learning structure that can extract highly
speaker-sensitive features. By employing this tool, we studied the SRE performance on three types of
trivial events: cough, laugh and ``Wei'' (a short Chinese ``Hello''). The results show that there is rich
speaker information within these trivial events, even for cough that is intuitively less speaker distinguishable.
With the deep feature approach, the EER can reach 10\%-14\% with the three trivial events, despite their
extremely short durations (0.2-1.0 seconds).

\end{abstract}

\section{Introduction}

Automatic speaker recognition (SRE) is an important biometric authentication technology. After several decades of development, SER has gained significant progress and the performance has been good enough for some constrained applications, e.g., with sufficient enrollment and test speech and the quality of the speech signals is reasonable~\cite{Kinnunen10,hansen2015speaker}.
In spite of the prominent success, however, almost all the existing SRE approaches work on long-duration linguistic speech segments, e.g., segments with clear and long linguistic content such as ``Hello, Google''~\cite{ehsan14}. In this paper, we are interested in some ``trivial events'' in speech signals, such as cough, laugh and ``Wei'' (a short ``hello'' in Chinese). These events are ubiquitous in conversations and often possess distinct properties, so it would be highly interesting to investigate how much speaker information is loaded in each of them. Moreover, SRE on trivial events may provide a powerful tool to attack speech disguising, as people who intend to disguise a personal identity are not easy to change her/his behaviors on these trivial events.

Very little has been done on these trivial events in SRE. This is not surprising: recognition on regular speech has been notoriously difficult, so it would be much harder to work with trivial events that are often short and may contain only non-linguistic content. Cough, for example, is often as short as 0.2 seconds, and the pronunciation is significantly different from regular speech: the airflow rushes out of the lung quickly and strongly vibrates the vocal cord. It is not easy to predict how much speaker information is involved in such a short and untypical signal, and it is even not easy to tell from our experience that if a speaker can be identified from a cough if we are not familiar with the person. Moreover, the training data for the trivial events are often very limited: although they are ubiquitous in conversations, the data volume takes only a very small proportion of the entire speech. We have not found any large-scale database that focuses on trivial events.

In this paper, we tackle the trivial event SRE problem by the deep speaker feature approach proposed recently by our group~\cite{li2017deep}. This approach designs a deep neural network (DNN) to learn frame-level speaker features
from vast raw data, where the input of the model is a frame plus a short context, and the target involves all the
speakers in the training database. The goal of the learning is to discover a speaker feature extractor (by the
DNN hierarchy) that can extract speaker-sensitive features from a window of speech frames. Li et al.~\cite{li2017deep}
reported very promising results and found that a good recognition accuracy can be obtained with even a very short speech segment.
This feature learning approach therefore provides a powerful tool to discover the speaker information load of a
small speech segment and retrieve it if it exists. We will use this tool to study the trivial events and try
to answer the following three questions:

\begin{itemize}
\item Does a particular trivial event involve speaker information?
\item Can the speaker information, if exists in a trivial event, be extracted from the event speech?
\item Can the deep feature model trained with a regular speech database be migrated to recognize trivial event segments?
\end{itemize}

\begin{figure}[htb]
\begin{center}
%\vspace*{50pt}
\includegraphics[width=0.95\linewidth]{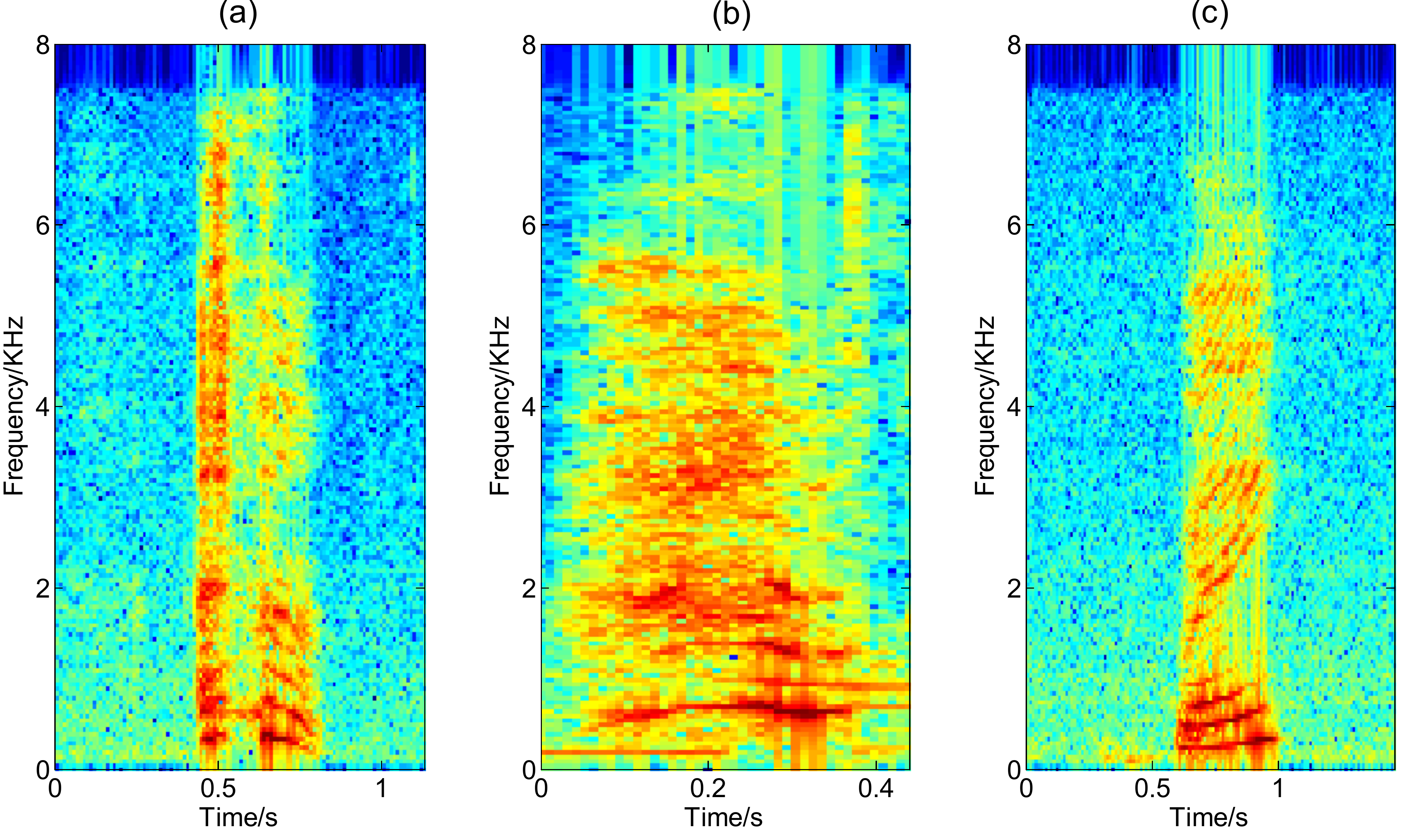}
\end{center}
\caption{Spectra of (a) Cough (b) Laugh (c) ``Wei'' spoken by the same person.}
\label{fig:spec}
\vspace*{-8pt}
\end{figure}

Our focus is put on three types of trivial events: cough, laugh and ``Wei''. We choose these types because they are among the most frequent ones in telephone conversations and are highly representative. First of all, cough is mostly related to vocal folder and involves little modulation from vocal track, so the spectrum contains no formants; laugh is also highly vocal-folder related, but the vocal tract may modulate the signal to some extent with the possible existed spectrum formant; ``Wei'' is mostly a regular speech segment, with a clear formant structure. Figure~\ref{fig:spec} displays the spectra of cough, laugh and ``Wei'' pronounced by the same speaker. It can be seen clearly that formant patterns exist in ``Wei'' but not clear in laugh, and totally absent in cough. From another perspective, cough is extremely short (less than 0.3 seconds), ``Wei'' is often very short (e.g., about 0.4 seconds), and laugh possesses a large variation in length, both intra-speaker and inter-speaker (e.g., from 0.3 to 1.0 seconds). Finally, the speaker information loads are intuitively different among these three types of events: it seems that ``Wei'' involves the most rich speaker information, as people can recognize who is speaking by a single ``Wei'' when picking up a phone. Laugh is the second and cough seems the mostly vague. In summary, investigation on the three types of events may provide a reasonable picture of the trivial event SRE.

The organization of this paper is as follows: we firstly describe the deep feature learning approach in brief in Section~\ref{sec:dfl}, and then present the trivial event speech database we collected for this study in Section~\ref{sec:data}. The experiments will be presented in Section~\ref{sec:exp}, followed by some conclusions and discussions in Section~\ref{sec:con}.

\section{Deep feature learning}
\label{sec:dfl}

Most existing successful SRE approaches are model-based. For example, the
famous Gaussian mixture model-universal background model (GMM-UBM) framework~\cite{Reynolds00}
and the subsequent subspace models, including the joint factor analysis approach~\cite{Kenny07}
and the i-vector model~\cite{dehak2011front}. They are generative models and heavily utilize
unsupervised learning. Improvements have been achieved by either discriminative compensation
(e.g., the SVM model~\cite{Campbell06} or PLDA~\cite{Ioffe06}) or subspace modeling~\cite{Kenny14,lei2014novel}.
Almost all these methods are based on raw acoustic features, e.g., the popular Mel frequency cepstral
coefficients (MFCC) feature.

An advantage of model-based approaches is that they can discover the group-based behavior of
speech signals of different speakers, and hence make appropriate decisions in the sense of maximum likelihood.
However, the heavy reliance on statically models largely impedes the passion of researchers in discovering
the inherent and essential mechanism that characterizes the traits of a speaker. Without knowing this
essence, existing methods have to rely on a long speech segment to identify a speaker, by observing
the `distributional patterns' of the speech frames within the segment.

Using the model-based approach to discriminate trivial event speech segments would be difficult, due to the limited
trivial event data for both training, enrollment and test. The only possible solution is to extract as much speaker information
as possible from the short and untypical trivial event speech, and use a model as simple as possible to make the discrimination.
This is the so called `feature-based' approach. Unfortunately, traditional feature-based methods rely on
human knowledge, which have been demonstrated to be ineffective even for regular speech, not to say
the trivial event speech for which our knowledge is far from rich.

Fortunately, our recent research shows that it
is possible to learn speaker sensitive features from raw speech signals by deep neural networks (DNN)~\cite{li2017deep}, inspired by
the seminar work of Ehsan~\cite{ehsan14}.
We found a simple DNN model that can learn speaker features very well, and a good SRE performance can be obtained
with a very small speech segment. This success in fact demonstrated that the speaker trait is
largely a short-term spectral property, rather than a long-term distribution pattern. It also offers
the possibility to discover the essential character of a speaker with very small trivial speech segments, for example, a
cough or laugh.

\begin{figure*}
    \centering
    \includegraphics[width=0.95\linewidth]{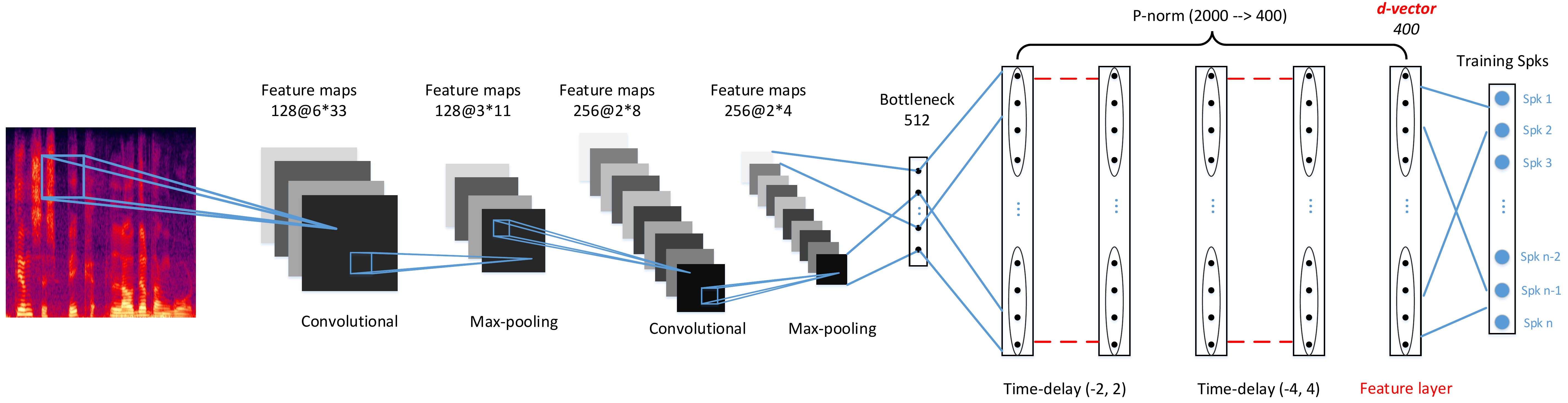}
    \caption{The CT-DNN structure used for deep speaker feature learning.}
    \label{fig:ctdnn}
    \end{figure*}

The DNN structure we designed involves a few convolutional layers and several time-delayed layers: the former
extracts local discriminative patterns,  while the latter allows a long temporal context. This is referred as
a CT-DNN model. Figure~\ref{fig:ctdnn} illustrates the CT-DNN structure used in this work.

More specifically, the CT-DNN structure consists of a convolutional (CN) component and a time-delay (TD)
component, connected by a bottleneck layer consisting of $512$ hidden units.
The convolutional component involves two CN layers, each followed by a max pooling.
The TD component involves two TD layers, each followed by a P-norm layer.
The settings for the two components, including the patch size, the number of feature maps, the
time-delay window, the group size of the P-norm, have been shown in Figure~\ref{fig:ctdnn}.
A simple calculation shows that with these settings, the size of the effective
context window is $20$ frames.
The output of the P-norm layer is projected to a feature layer consisting of $400$ units,
which is connected to the output layer whose units correspond to the speakers in the training data.

This CT-DNN model can be trained easily by the natural stochastic gradient
descent (NSGD)~\cite{povey2014parallel} algorithm.
Once it has been trained, the speaker feature can be read from the feature layer,
i.e., the last hidden layer of the model. As in~\cite{ehsan14}, the utterance-level representation, called a `d-vector',
is derived by simply averaging the speaker features of all the frames within the utterance.

During test, the d-vectors of the enrollment and test utterances are produced respectively.
The cosine distance between these two vectors is then used as the decision score for the SRE task.
Similar to the i-vector system, some simple normalization
methods can be employed to enhance the SRE performance, such as linear discriminant analysis (LDA)
and probabilistic LDA (PLDA).

When applying the above deep feature learning approach to trivial event SRE, a particular problem is that
the training data is highly sparse. It would be difficult to collect a large amount of trivial event speech
segments, particularly cough and laugh. In this study, we simply employ a model that was trained with a large
regular speech database. Our assumption is that the training data, although not aiming for trivial events,
still involve some valuable information for them; particularly, some trivial event speech may be represented by
regular phones, e.g., ``Wei''. This setting will also test the generalization capability of the
deep feature learning model, and examine whether the speaker characterization produced from the trivial event speech is the same as that from the regular speech.

\section{Database construction}
\label{sec:data}

Although the model can be trained with a regular speech database, the test data has to be
specifically prepared. Since there is not any public database suitable for our study,
we decided to construct a small trivial event speech database for the test, and release
the data for public usage. This database is
denoted by `CSLT-COUGH100', and can be download online\footnote{http://data.cslt.org}.
Table~\ref{tab:cough100} presents the data profile in details.

\begin{table}[htp]
    \begin{center}
        \caption{Data profile of CSLT-COUGH100}
        \label{tab:cough100}
          \begin{tabular}{|l|c|c|c|c|}
           \hline
                          &   Spks    & Total Utts  &   Utts/Spk   &   Avg. dur   \\
           \hline
               Cough      &   104     &    890      &    8.6       &     0.27s     \\

           \hline
               Laugh      &   104     &   904      &     8.7      &     0.33s   \\

           \hline
               ``Wei"        &   104     &   848       &    8.2     &      0.37s  \\
           \hline
          \end{tabular}
    \end{center}
\end{table}

To collect the data, we designed a simple Android application, which instructs users to click a button
to record cough, laugh and ``Wei''.
The recording involved three sessions, each for one type of event. In each session, the speaker was
instructed to pronounce the requested event (cough, laugh, ``Wei'') multiple times (not less than 8 times), with
any variation they preferred. The recordings were then segmented into short segments by hand, each containing only
a single event. The sampling rate of the recording is 16k Hz and the precision of the samples is 16 bits. The recording is mostly
in the office environment, but some of them are collected on the street. The age of the participants ranges from
$20$ to $60$, although most of them are between $20$-$30$.

\section{Experiments}
\label{sec:exp}

This section reports our experimental results. We first describe the data and settings, and then report the
SRE results in terms of the equal error rate (EER). Some analysis for the discriminative power of the deep features
are also presented.

\subsection{Data}

The Fisher database was used as the training set, which is recorded by telephone and the sampling rate
is 8k Hz. The CSLT-COUGH100 was used as the test set. As the origin data of CSLT-COUGH100 is in 16k Hz, we
down-sampled the signals to 8k Hz to match the Fisher database. More details of the two datasets are as follows.

\begin{itemize}
    \item \textbf{Training set}: It consists of $2,500$ male and $2,500$ female speakers, with $95,167$ utterances randomly selected from the \emph{Fisher} database, and each speaker has about $120$ seconds of speech segments.
        This data set was used to train the UBM, the T matrix, and the LDA/PLDA models of the i-vector system, as well as the CT-DNN model of the d-vector system.
    \item \textbf{Test set}: The CLST-COUGH100 database, consisting of $104$ speakers. The database contains three types of trivial events (cough, laugh and ``Wei" ), and each type of event involves $5$-$10$ segments. Details have been shown in Table~\ref{tab:cough100}.

\end{itemize}

\subsection{Model settings}

For the purpose of comparison, an i-vector system was constructed as the baseline system.
The raw features of this system involve $19$-dimensional MFCCs plus the log energy.
This raw features were augmented by the first and second order derivatives, resulting in
feature vectors of 60 dimensions.
The UBM was composed of $2,048$ Gaussian components, and the dimensionality of
the i-vector space was $400$. The dimensionality of the LDA projection space
was set to $150$. When using PLDA as the scoring metric, the i-vectors were
length normalized.
The system was trained using the Kaldi SRE08 recipe~\cite{povey2011kaldi}.

The d-vector system uses the CT-DNN architecture shown in Figure~\ref{fig:ctdnn}.
The input features were 40-dimensional Fbanks. A symmetric $4$-frame window was used to
splice the neighboring frames, resulting in $9$ frames in total.
The number of output units was $5,000$, corresponding to the number of speakers in
the training data.
The frame-level speaker features were extracted from the last hidden layer (the feature layer in Figure~\ref{fig:ctdnn}),
and the utterance-level d-vectors were derived by averaging the frame-level speaker features.
The scoring methods used for the i-vector system were also used for the
d-vector system during the test, including cosine distance, LDA and PLDA.

\subsection{Main results}

The EER results of the i-vector system and the d-vector system are reported in Table~\ref{tab:baseline}.
It can be observed that with the best i-vector baseline (cosine scoring), the performance on ``Wei'' (12.72\%) is
reasonably good considering the short duration of the test utterance. On cough and laugh, the
performance is significantly reduced (19.96\% and  23.03\% respectively). These results are expected,
as the model are not intentionally trained to cover these two kinds of trivial events, and
the content of these two events are largely non-linguistic, so likely involve less speaker information.

    \begin{table}[htb]
    \begin{center}
      \caption{EER(\%) results with the i-vector and d-vector systems.}
      \label{tab:baseline}
          \begin{tabular}{|l|l|c|c|c|c|}
            \hline
            \multicolumn{2}{|c|}{}                 &\multicolumn{3}{c|}{EER\%}\\
            \hline
               Systems              &  Metric    &   Cough &    Laugh   &  ``Wei"  \\
            \hline
               i-vector             &    Cosine   &    19.96   &    23.03      &  12.72           \\
                                    &    LDA      &    23.55   &    24.24      &  12.90            \\
                                    &    PLDA     &    23.33   &    24.30      &  13.77            \\
           \hline
               d-vector             &    Cosine   &     11.19  &   13.62       &  10.66          \\
                                    &    LDA      &     12.37  &   13.41       &  10.75          \\
                                    &    PLDA     &     10.99  &   13.76       &  10.06        \\
           \hline
          \end{tabular}
      \end{center}
   \end{table}

The d-vector system is significantly better than the i-vector system,
demonstrating that the feature-based approach is more powerful.
The best d-vector system is the one with the PLDA scoring, and the EERs with this system can reach
$10.99\%$ ,$13.76\%$ and $10.06\%$ for cough, laugh and ``Wei'', respectively. The lower EER for ``Wei''
compared to the i-vector system is something expected, as we have demonstrated in~\cite{li2017deep} that the d-vector
system is stronger than the i-vector system on small speech segments such as ``Wei''.
The good performance with cough and laugh, however, is a bit surprising: both the two events do not contain
linguistic content, but the performance is not significantly worse than the linguistic event ``Wei''. 
This seems that non-linguistic events still involve rich speaker information, and implies that
our vocal cords are highly complex and speaker specific.
The low performance with the i-vector model on
the trivial events should not be explained as little speaker information embedded within these events; instead, it
is just because the i-vector model cannot extract the information and utilize it well.

\begin{figure}[htb]
\begin{center}
%\vspace*{20pt}
\includegraphics[width=1\linewidth]{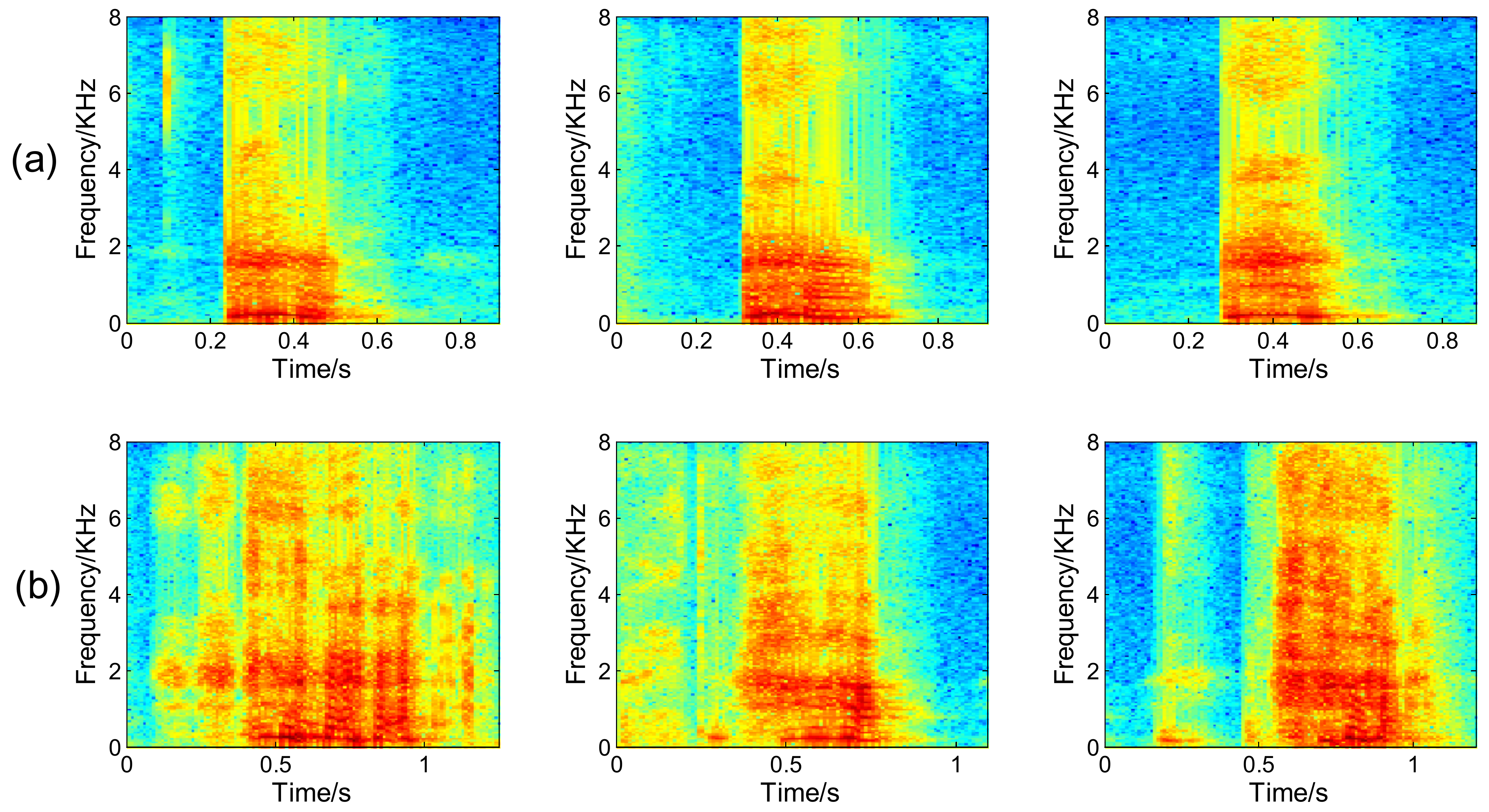}
\end{center}
\caption{The spectra of (a) Cough (b) Laugh segments of the same speaker.}
\label{fig:compare}
%\vspace*{-8pt}
\end{figure}

Comparing the results on cough and laugh, it can be seen that the performance on laugh is slightly worse than
on cough. This is again a little unexpected.
From our experience, it seems that we can recognize a person easier by her/his laugh than cough.
A possible explanation is that the laugh speech may involve significant within-speaker variations, due to the
freedom within the vocal tract modulation. The cough, on the other hand, is less modulated by the vocal tract
and thus more stable.
To have an intuitive comparison, Figure~\ref{fig:compare} shows the spectra of three cough segments and three laugh
segments from the same speaker. It can be seen that the three laugh segments are clearly different,
while the cough segments are pretty much the same.

Finally, we found that for both two systems, the discriminative normalization approaches, LDA and PLDA, did not
provide clear advantage. This is particularly the case for the i-vector system for which all the
normalization methods reduce the performance on all the three types of trivial events.
A possible reason is that trivial events are different from regular speech, so the
LDA and PLDA models trained with the regular speech database are not very suitable.
This argument explains why the performance reduction with the normalization methods
is most significant on cough, if we admit that cough is the most distinct from regular speech
among the three types of events.

\subsection{Feature discrimination}

\begin{figure*}
\centering
\subfigure[Cough]{
\label{fig:subfig:a}
%% label for first subfigure
\includegraphics[width=0.3\linewidth]{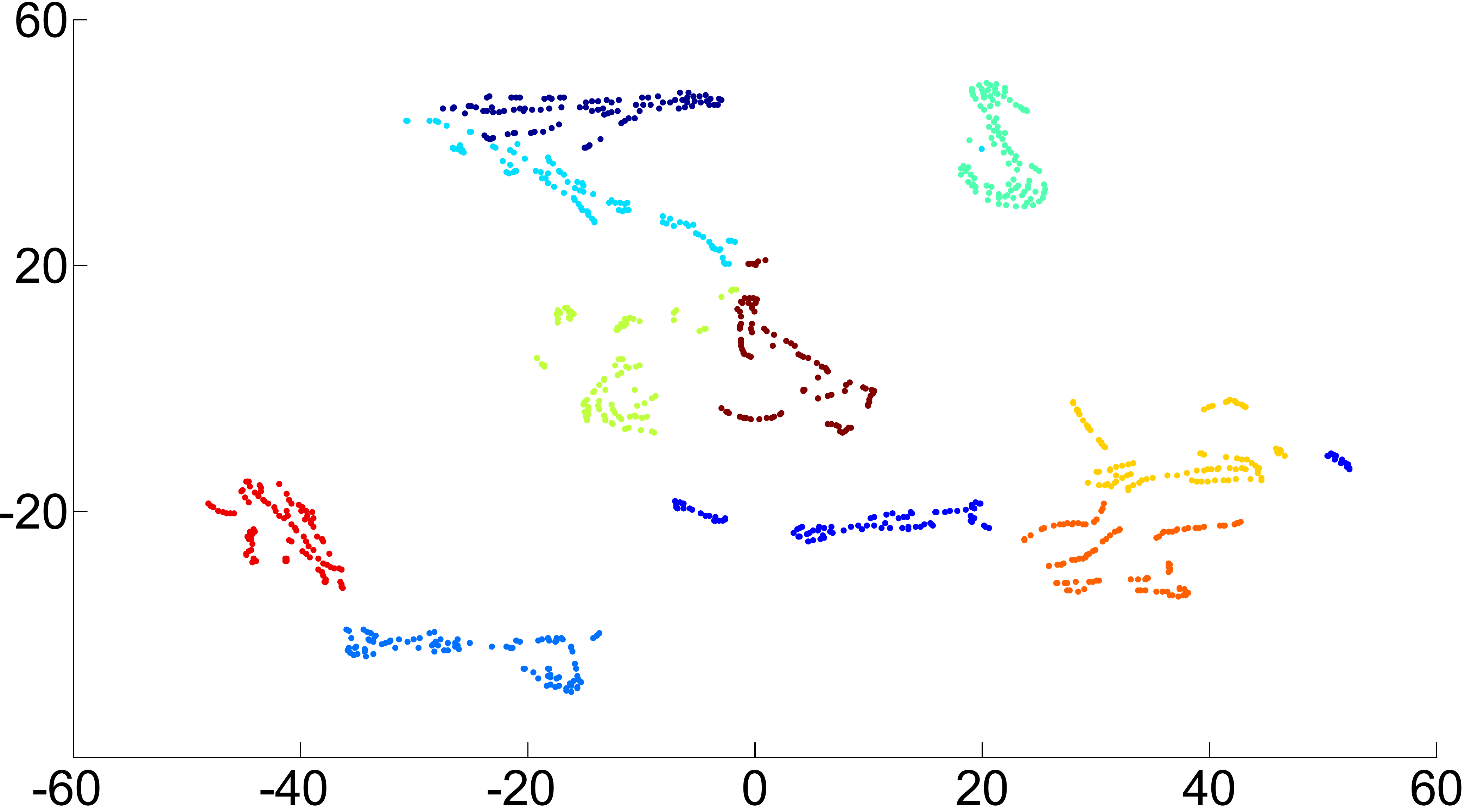}
}
\hspace{0in}
\subfigure[Laugh]{
\label{fig:subfig:b}
%% label for second subfigure
\includegraphics[width=0.3\linewidth]{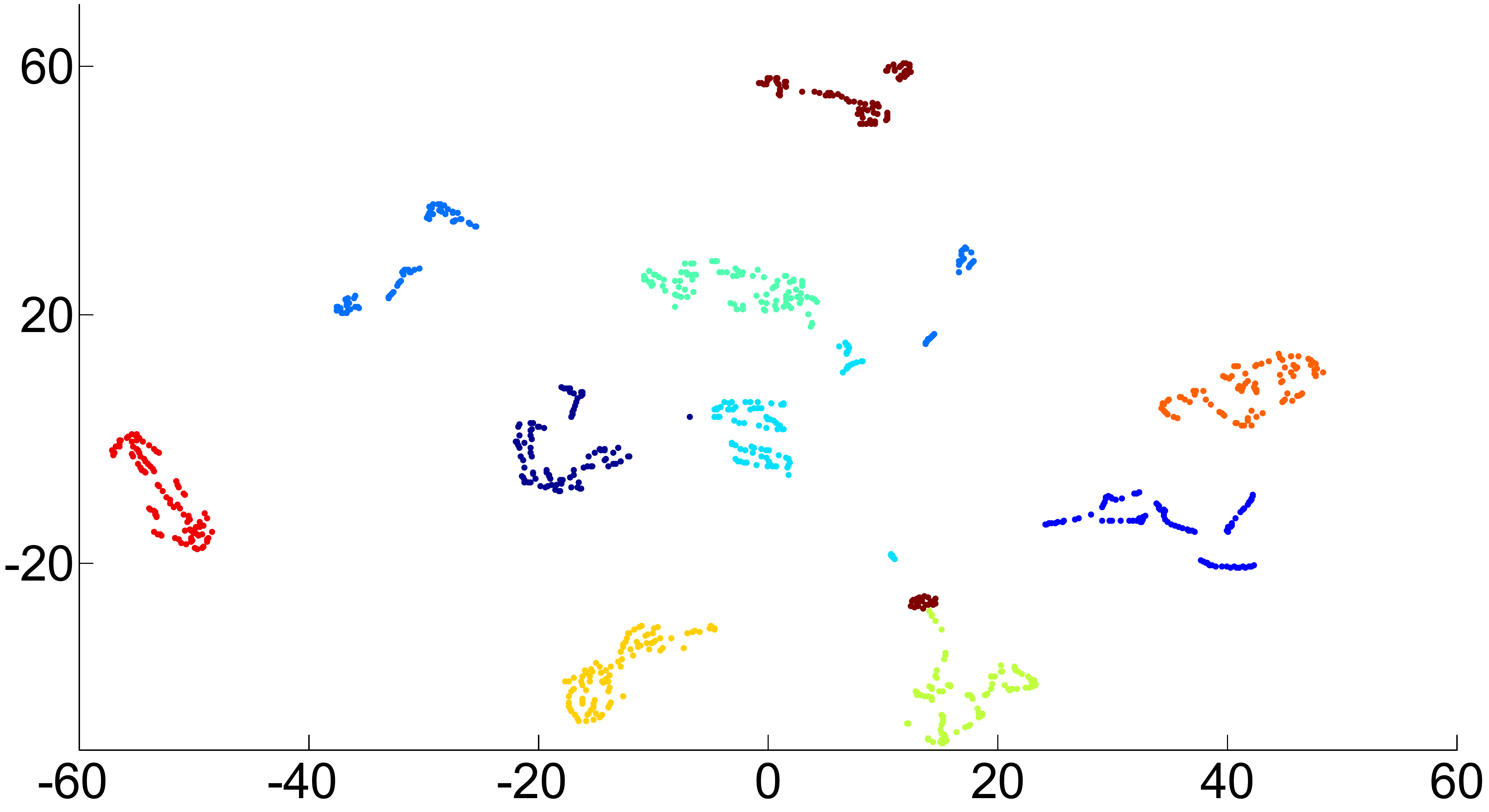}
}
\hspace{0in}
\subfigure[Wei]{
\label{fig:subfig:c}
%% label for second subfigure
\includegraphics[width=0.3\linewidth]{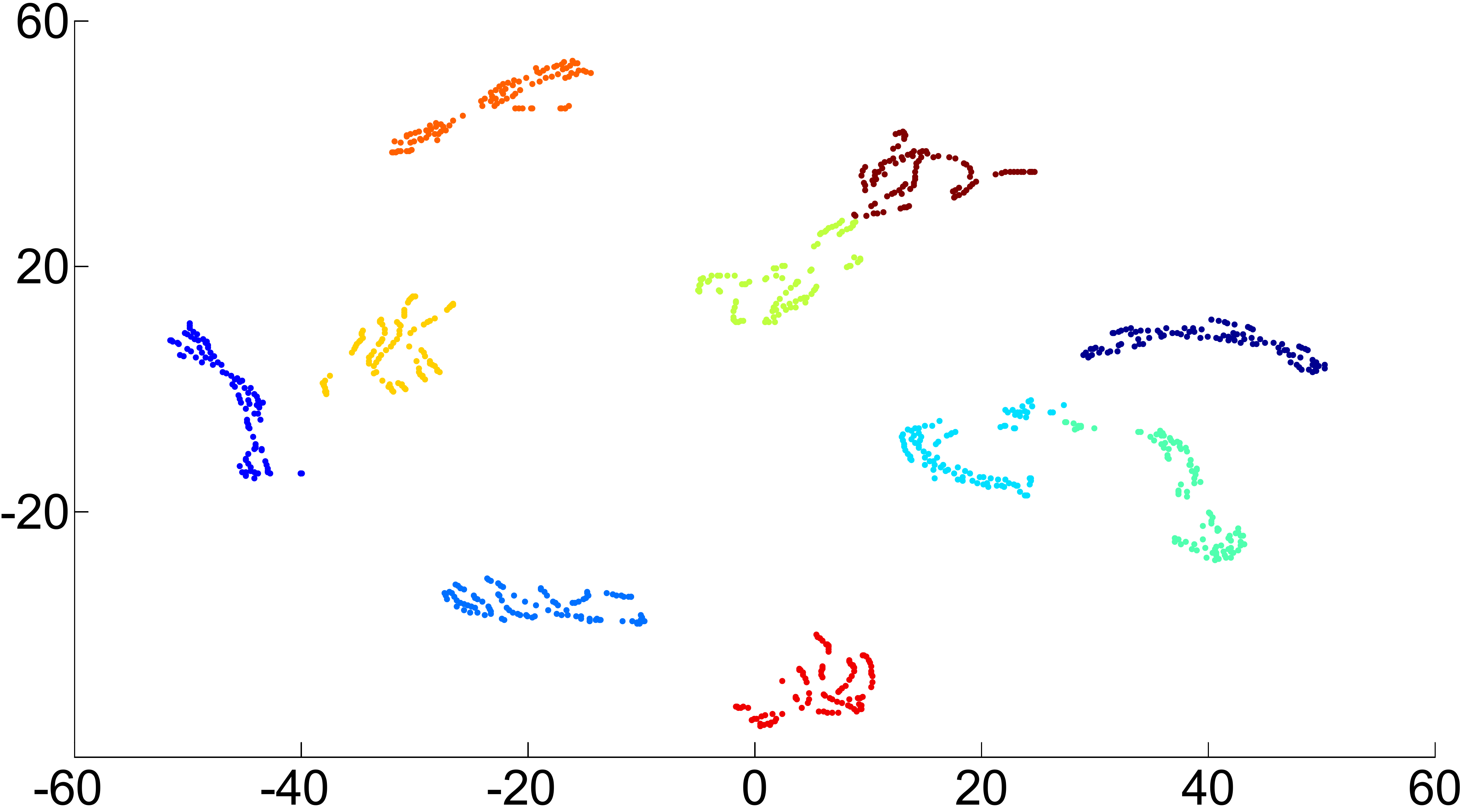}
}
\caption{Deep speaker features on events (a) Cough (b) Laugh (c) ``Wei'' randomly sampled from 10 speakers. The pictures are plotted by t-SNE, with each color representing a speaker. }
\label{fig:subfig} %% label for entire figure
\end{figure*}

In the last experiment, we investigate the distribution of the deep speaker features, particularly
the speaker variations with these three types of trivial events. For this purpose, we randomly
selected $10$ speakers, and drew the speaker features belonging to these three types of events respectively using t-SNE~\cite{saaten2008}. The results are presented in Figure~\ref{fig:subfig}. It can be seen that the learned features with ``Wei" are reasonably discriminative for speakers. But there are still variations appeared in cough and laugh figures as seen from the plot (a) and (b).

\section{Conclusions}
\label{sec:con}

This paper employed the deep speaker feature learning approach to perform SRE on trivial events, and the focus was put on cough, laugh and ``Wei". Our experiments showed that the deep feature model trained on a regular speech database (Fisher) can be used to perform trivial event SRE with an unexpected success. In spite of the extremely short duration, the EER can be as low as 10\%-14\%, depending on the type of events. These results can answer the questions raised in the introduction session:

\begin{itemize}
\item Does a particular trivial event involve speaker information? Yes. At least for the three trivial events studied in this paper, rich speaker information is involved.
\item Can the speaker information, if exists in a trivial event, be extracted from the short segment? Yes. The deep feature approach was capable of extracting the speaker information from the short and idiocratic trivial events.
\item Can the deep feature model trained with a regular speech database be migrated to recognize trivial event segments? Yes. A DNN model trained with the Fisher database worked well on trivial event SRE.
\end{itemize}

There is much work remaining: how about the performance on other trivial events, e.g., En, Ah, tapping and flapping speech? What is the implication of the experimental results for the acoustic and linguistic research? How the performance will be in a true scenario of speech disguise? All are under investigation.

%\section*{Acknowledgment}

%\newpage
\bibliographystyle{IEEEtran}
\bibliography{mybib}
\end{document}